\documentclass[12pt]{article}

\catcode`\@=11
\def\marginnote#1{}
\newcount\hour
\newcount\minute
\newtoks\amorpm
\hour=\time\divide\hour by60
\minute=\time{\multiply\hour by60 \global\advance\minute by-\hour}
\edef\standardtime{{\ifnum\hour<12 \global\amorpm={am}%
        \else\global\amorpm={pm}\advance\hour by-12 \fi
        \ifnum\hour=0 \hour=12 \fi
        \number\hour:\ifnum\minute<10 0\fi\number\minute\the\amorpm}}
\edef\militarytime{\number\hour:\ifnum\minute<10 0\fi\number\minute}
\def\draftlabel#1{{\@bsphack\if@filesw {\let\thepage\relax
   \xdef\@gtempa{\write\@auxout{\string
      \newlabel{#1}{{\@currentlabel}{\thepage}}}}}\@gtempa
   \if@nobreak \ifvmode\nobreak\fi\fi\fi\@esphack}
        \gdef\@eqnlabel{#1}}
\def\@eqnlabel{}
\def\@vacuum{}
\def\draftmarginnote#1{\marginpar{\raggedright\scriptsize\tt#1}}
\def\draft{\oddsidemargin -.5truein
        \def\@oddfoot{\sl preliminary draft \hfil
        \rm\thepage\hfil\sl\today\quad\mil922 itarytime}
        \let\@evenfoot\@oddfoot \overfullrule 3pt
        \let\label=\draftlabel
        \let\marginnote=\draftmarginnote
   \def\@eqnnum{(\theequation)\rlap{\kern\marginparsep\tt\@eqnlabel}%
\global\let\@eqnlabel\@vacuum}  }

\def\preprint{\twocolumn\sloppy\flushbottom\parindent 1em
        \leftmargini 2em\leftmarginv .5em\leftmarginvi .5em
        \oddsidemargin -.5in    \evensidemargin -.5in
        \columnsep 15mm \footheight 0pt
        \textwidth 250mmin      \topmargin  -.4in
        \headheight 12pt \topskip .4in
        \textheight 175mm
        \footskip 0pt
        \def\@oddhead{\thepage\hfil\addtocounter{page}{1}\thepage}
        \let\@evenhead\@oddhead \def\@oddfoot{} \def\@evenfoot{} }

\def\titlepage{\@restonecolfalse\if@twocolumn\@restonecoltrue\onecolumn
     \else \newpage \fi \thispagestyle{empty}\c@page\z@ 
        \def\thefootnote{\fnsymbol{footnote}} }

\def\endtitlepage{\if@restonecol\twocolumn \else  \fi
        \def\thefootnote{\arabic{footnote}} \setcounter{footnote}{0}}

\catcode`@=12
\relax

\def\bea{\begin{array}}
\def\bem{\begin{displaymath}}
\def\beq{\begin{equation}}

\def\eea{\end{array}}
\def\eem{\end{displaymath}}
\def\eeq{\end{equation}}

\def\s2w{\sin^2 \theta_W}

\relax
\def\be{\begin{equation}}
\def\ee{\end{equation}}
\def\ba{\begin{eqnarray}}
\def\ea{\end{eqnarray}}

\def\w{\wedge}
\def\d{{\rm d}}

\def\r{\rho}
\def\a{\alpha}

\def\b{\beta}

\def\g{\gamma}

\def\dd{\delta}

\def\e{\epsilon}

\def\p{\pi}

\def\m{\mu}
\def\n{\nu}

\def\l{\lambda}

\def\s{\sigma}

\def\IR{\relax{\rm I\kern-.18em R}}

\def\inv{^{\raise.15ex\hbox{${\scriptscriptstyle -}$}\kern-.05em 1}}

\def\be{\begin{equation}}
\def\ee{\end{equation}}
\def\ba{\begin{eqnarray}}
\def\ea{\end{eqnarray}}

\def\a{\alpha}
\def\b{\beta}
\def\g{\gamma}

\def\d{{\rm d}}
\def\e{\epsilon}
\def\p{\psi}

\def\m{\mu}
\def\n{\nu}
\def\r{\rho}
\def\l{\lambda}

\def\s{\sigma}
\def\o{\omega}
\def\f{\phi}

\def\ks{{k \kern-.5em /}}
\def\es{{\e \kern-.4em /}}
\def\ds{{\partial \kern-.5em /}}
\def\Ds{{D \kern-.6em /}}

\def\R{{\cal R}}

\def\eb{{\overline e}}

\def\inv{^{\raise.15ex\hbox{${\scriptscriptstyle -}$}\kern-.05em 1}}

\def\wt{\widetilde}

\relax

\renewcommand{\theequation}{\thesection.\arabic{equation}}
\begin{document}
\topmargin-1.0cm
%
%
%
%
\def\F{\Phi}
\def\la{\langle}
\def\ra{\rangle}
\def\st{{}^*}
\def\et{\tilde e}
\def\sty{\, {}^{*_Y}\kern-.2em}
\def\rh{\hat r}
\begin{titlepage}
\begin{flushright}
LPTENS-02/61\\ hep-th/0302021 \\ February 2003
\end{flushright}
\vskip 3.5cm

\begin{center}{\Large\bf Compact weak $G_2$-manifolds with 
conical singularities }
\vskip 1.5cm
{\bf Adel Bilal$^{1}$ and Steffen Metzger$^{1,2}$
}
\vskip.3cm
$^1$ CNRS - Laboratoire de Physique Th\'eorique, 
\'Ecole Normale Sup\'erieure\\
24 rue Lhomond, 75231 Paris Cedex 05, France

\vskip.3cm
$^2$ Sektion Physik, Ludwig-Maximilians-Universit\"at\\ 
Munich,
Germany\\

\vskip.3cm
{\small e-mail: {\tt adel.bilal@lpt.ens.fr, 
metzger@physique.ens.fr}}
\end{center}
\vskip .5cm

\begin{center}
{\bf Abstract}
\end{center}
\begin{quote}
We construct 7-dimensional compact Einstein 
spaces with conical singularities that preserve 1/8
of the supersymmetries of M-theory. Mathematically they have 
weak $G_2$-holonomy. We show that for every non-compact 
$G_2$-holonomy manifold which is asymptotic to a cone on 
a 6-manifold $Y$, there is a corresponding weak 
$G_2$-manifold with two conical singularities which, close to 
the singularities, looks like a cone on $Y$. Our construction
provides explicit metrics on these weak $G_2$-manifolds.
We completely determine the cohomology of these manifolds
in terms of the cohomology of $Y$.

\end{quote}


\end{titlepage}
\setcounter{footnote}{0}
\setcounter{page}{0}
\setlength{\baselineskip}{.7cm}
\newpage
%
%

\section{Introduction\label{Intro}}

When discussing compactifications of M-theory on a 7-manifold $X$
down to 4 dimensions one is mainly interested in preserving
exactly ${\cal N}=1$ supersymmetry. If the 4-form field $G$ of
11-dimensional supergravity has a vanishing expectation value, it
is well-known that the compactification manifold $X$ must be
Ricci-flat with holonomy group  $G_2$. This can however be
generalized, similarly to the Freund-Rubin solution, to a
non-vanishing background value with $G_{\m\n\r\s}$ being
proportional to the 4-dimensional $\e$-tensor. In this latter
case the compactification manifold must have so-called weak
$G_2$-holonomy if we want to preserve exactly ${\cal N}=1$
supersymmetry. Weak $G_2$-holonomy manifolds are Einstein spaces
with positive curvature, in agreement with the fact that the
non-vanishing 4-form field induces a non-vanishing
energy-momentum tensor.

M-theory on $G_2$-holonomy manifolds has been much discussed
recently and many such manifolds are known with more or less
explicit metrics. Complete non-compact metrics were first given
in \cite{BS}, and compact spaces, though no explicit metrics,
were first constructed in \cite{Joyce1}, see also \cite{Joyce2}.

M-theory/supergravity compactified on smooth $G_2$-holonomy
manifolds only has abelian gauge groups and no charged chiral
fermions, and hence is rather uninteresting. Introducing
$ADE$-orbifold singularities, however, leads to non-abelian gauge
groups, the symmetry enhancement being provided by M2-branes that
wrap the vanishing two-cycles \cite{ADE}. The presence of conical
singularities was shown to lead to charged chiral fermions
\cite{Cvetic,AW,AchW}, and hence the theory could be 
potentially anomalous.
The issue of anomaly cancellation is discussed in
\cite{Witten,BM}. The drawback of these discussions is that no
explicit examples of {\it compact} $G_2$-holonomy manifolds with
conical singularities are known. Instead one considers \cite{AW}
the non-compact manifolds of \cite{BS} which in a limit become
cones on some 6-manifold $Y$. One then {\it assumes} that compact
$G_2$-holonomy manifolds also can develop conical singularities
and, close enough to the singularities, look like one of these
cones. While these discussions are very elegant, it still would
be nice to have  some explicit examples of compact manifolds with
conical singularities at our disposal.

Since the basic examples of conical singularities are limits of
the non-compact manifolds of \cite{BS} one could try to start
with these manifolds and somehow make them compact. This can
indeed be done, as we will show in this paper,
 at the price of introducing positive
curvature, deforming the $G_2$-holonomy to weak $G_2$-holonomy.

Our strategy to construct the compact weak $G_2$-holonomy
manifolds is the following: we begin with any non-compact
$G_2$-holonomy manifold $X$ that asymptotically, for ``large
$r$'' becomes a cone on some 6-manifold $Y$. The $G_2$-holonomy
of $X$ implies certain properties of the 6-manifold $Y$ which we
deduce. In fact, $Y$ can be any Einstein space of positive
curvature with weak $SU(3)$-holonomy. Then we use this $Y$ to
construct a compact weak $G_2$-holonomy manifold $X_\l$ with two
conical singularities that, close to the singularities, looks
like a cone on $Y$.

We go on to study in detail the cohomology of these manifolds.
Since, up to scale, $X_\l$ is completely determined in terms of
$Y$ it is not surprising that the cohomology of $X_\l$ is
determined by that of $Y$. Due to the singularities, however, one
has to specify which class of forms one is going to allow on
$X_\l$. Physically it is clear that one is interested in
square-integrable forms. We prove that {\it all} $L^2$-harmonic
$p$-forms on $X_\l$ for $p\le 3$ are given by the trivial
extensions of the harmonic $p$-forms on $Y$. In particular,
$b^2(X_\l)=b^2(Y)$ and  $b^3(X_\l)=b^3(Y)$, while $b^1(X_\l)=0$.
For $p\ge 4$ the  $L^2$-harmonic $p$-forms on $X_\l$ are just the
Hodge duals of the previous ones. We also give a simple
generalisation of these cohomological results to analogous
constructions in arbitrary dimensions of spaces with two conical
singularities.

Our construction provides examples, with explicitly known metrics
and cohomology, of compact manifolds with conical singularities
that are Einstein spaces and preserve ${1\over 8}$ of the
supersymmetries of M-theory. Actually, our results can easily
be extended and applied more widely.
One could go on and further quotient
$X_\l$ by some $\Gamma_{ADE}$ to obtain $ADE$-orbifold
singularities. Almost the whole discussion of 
\cite{AW,AchW,Witten,BM}
about non-abelian gauge groups, chiral fermions and anomalies
could then be repeated in this setting, but now with the 
advantage of having well-defined explicit examples at hand.
Other situations for which our results yield compact examples 
are those discussed in refs. \cite{AMV,Fried}.

This paper is organized as follows. Section 2 is a brief review
of $G_2$- and weak $G_2$-holonomy. In section 3 we explicitly
carry out the construction of the weak $G_2$-holonomy manifolds,
and in section 4 we discuss the cohomology of these manifolds in
detail and prove the above-mentioned results. In section 5, we
draw some conclusions and discuss further developments. The
appendix contains some technical material which is needed in the
main text.

\section{A brief review of weak $G_2$-holonomy}
\setcounter{equation}{0}

Compactifications of M-theory on 7-manifolds $X$ of 
$G_2$-holonomy preserve ${1\over 8}$ of the 
32 supersymmetries if
the expectation value of the four-form $G$ vanishes. It is 
known, however, that one can have non-vanishing $G$-flux on the 
four-dimensional space-time $M_4$ and still preserve the 
same amount of supersymmetry if the 7-manifold $X_\l$ has 
{\it weak} $G_2$-holonomy and $M_4$ is $AdS_4$. There are
several equivalent ways to characterise these manifolds. 
For $G_2$-holonomy we have (exactly) one covariantly constant 
spinor $\eta$
\be\label{f1}
\left( \partial_j+{1\over 4} \o_j^{ab} \g^{ab}\right) \eta =0 \ ,
\ee
from which one can construct (see appendix \ref{appsd}) 
a closed and co-closed three-form 
$\F$:
\be\label{f2}
\d\F=0 \quad , \quad \d \st\F=0 \ .
\ee
Here $\o^{ab}$ is the spin-connection 1-form on the 7-manifold,
and $a,b=1,\ldots 7$ are flat indices, while $i,j=1, \ldots 7$
are curved ones.
This implies that $X$ is Ricci flat. 

For weak $G_2$ we have instead
\be\label{f3}
\left( \partial_j+{1\over 4} \o_j^{ab} \g^{ab}\right) \eta 
= i {\l\over 2} \g_j \eta\ ,\qquad \l\in {\bf R} \ ,
\ee
which implies the existence of a three-form $\F_\l$ obeying
\be\label{f4}
\d\F_\l = 4 \l\  \st\F_\l \ ,
\ee
as well as $X_\l$ being Einstein:
\be\label{f5}
\R_{ij}=6\l^2 g_{ij} \ .
\ee
It can be shown that the converse statements are also true, 
namely that eq. (\ref{f2}) implies  (\ref{f1}), and 
eq. (\ref{f4}) implies  (\ref{f3}). Note that for 
$\l\to 0$, at least formally, weak $G_2$ goes over to 
$G_2$-holonomy.

Physically, a non-vanishing Ricci tensor is due to a 
non-vanishing energy-momentum tensor $T_{MN}$. Indeed, 
eq. (\ref{f3}) precisely is the condition (see e.g. 
\cite{BDS}\footnote{
In \cite{BDS} $\l$ was normalised differently: 
$\l_{\rm BDS} =4\l$.})
for preserving ${\cal N}=1$ supersymmetry in four dimensions 
if the four-form $G$ has a background value
\be\label{f6}
\la G\ra =-6\l\  {\rm vol}_4
\ee
where $ {\rm vol}_4$ is the volume form on $M_4$, i.e. 
$\la G_{\m\n\r\s}\ra=-6\l\e_{\m\n\r\s}$ and all other 
components vanish. This induces a non-vanishing 
energy-momentum tensor, $T_{\m\n}\ne 0$ and $T_{ij}\ne 0$. 

\noindent
Einstein's equations are
\be\label{f7}
\R_{MN}-{1\over 2} g_{MN} \R = {1\over 96} 
\left( 8 G_{MPQR}G_N^{\ \ PQR} - g_{MN} G_{PQRS}G^{PQRS}\right)
\ee
and, for the background (\ref{f6}), they imply
\be\label{f8}
\la \R_{\m\n}\ra = -12\l^2 \la g_{\m\n}\ra \quad , \quad 
\la \R_{ij }\ra = 6 \l^2 \la g_{ij }\ra  \ .
\ee
This is consistent with eq. (\ref{f5}) and the fact that 
$M_4$ is $AdS_4$.

\section{Construction of weak $G_2$-holonomy manifolds with 
singularities}
\setcounter{equation}{0}

One method to construct  $G_2$-holonomy or weak $G_2$-holonomy
metrics is based on a certain 
self-duality condition for the spin connection, as explained 
in \cite{BDS}. This condition can be written as
\be\label{f9}
\psi_{abc} \o^{bc} =-2\l \, e^a
\ee
where $e^a$ is the 7-bein on $X$ and the $\psi_{abc}$ are 
the structure constants of the imaginary octonions.\footnote{
Note that $a,b,c$ are ``flat'' indices with euclidean signature,
and upper and lower indices are equivalent.
}
The latter are completely antisymmetric and equal 
$\pm 1$ or 0. An explicit choice is
$\psi_{123} = \psi_{516} = \psi_{624} = \psi_{435} = \psi_{471} 
= \psi_{673} = \psi_{572} = 1$. Self-dual and anti self-dual
projections are explained in appendix \ref{appproj}. It was 
shown in \cite{BDS} that $G_2$-holonomy/weak $G_2$-holonomy 
is equivalent to the existence of a local $SO(7)$ frame 
where the corresponding $\o^{bc}$ satisfy eq. (\ref{f9}).
We will call such a frame a self-dual frame. Most examples of 
$G_2$ metrics ($\l=0$) known in the 
literature actually are written in a self-dual frame. 
Examples of weak $G_2$-metrics 
naturally written with self-dual frames
are those with principal orbits being the Aloff-Walach 
spaces $SU(3)/U(1)_{k,l}$ as given in \cite{Kano},
as well as the one described in \cite{BDS}. These examples 
are smooth manifolds. They can be obtained by starting 
with an 8-dimensional manifold of $Spin(7)$-holonomy 
that is a cone over a 7-manifold $X$. An analogous 
self-duality condition for the 8-manifold then reduces 
to the self-duality condition (\ref{f9}) of $X$ with 
$\l\ne 0$ which has weak $G_2$-holonomy. In this 
construction the original $Spin(7)$ metric is a 
cohomogeneity-one metric so that the resulting 
metric on $X$ is cohomogeneity-zero.

A basic role in establishing $G_2$- or weak $G_2$-holonomy 
is played by the above-mentioned 3-form $\F$ or $\F_\l$.
Hence, it is worthwhile to note the following result.
The 3-form $\F$ of eq. (\ref{f2}) or $\F_\l$ of
(\ref{f4}) is given by\footnote{
Here we want to treat both cases in parallel and we simply write 
$\F_\l$ with the understanding that $\F_\l\Big\vert_{\l=0}=\F$.}
\be\label{f9a}
\F_\l={1\over 6} \psi_{abc}\ e^a\w e^b\w e^c
\ee 
if and only if the 7-beins $e^a$ are a self-dual frame. 
The proof is simple and is given in appendix \ref{appsd}.
Later, for weak $G_2$,
we will consider a frame which is not self-dual 
and thus the 3-form $\Phi_\l$ will be slightly more 
complicated than (\ref{f9a}).

Our goal is to construct compact weak $G_2$-holonomy manifolds 
with conical singularities.  Some specific examples of such
manifolds were constructed in \cite{Cleyton}.
Here we want to extend this construction. We will show that, 
for every non-compact $G_2$-manifold that is asymptotic 
to a cone on $Y$, one can construct a corresponding compact 
weak $G_2$-manifold with conical singularities, that close 
to each of the singularities becomes a cone on $Y$.

We start with any $G_2$-manifold $X$ which asymptotically 
is a cone on a compact 6-manifold $Y$:
\be\label{f10}
\d s_X^2 \sim \d r^2 +r^2 \d s_Y^2 \ .
\ee
Since $X$ is Ricci flat, $Y$ must be an Einstein manifold with 
$\R_{\a\b}=5\delta_{\a\b}$. In practice \cite{BS}, 
$Y={\bf CP}^3,\ S^3\times S^3$
or $SU(3)/U(1)^2$, with explicitly known metrics.  
On $Y$ we introduce 6-beins
\be\label{f11}
\d s_Y^2 = \sum_{\a=1}^6 \et^\a \otimes \et^\a \ ,
\ee
and similarly on $X$
\be\label{f12}
\d s_X^2 = \sum_{a=1}^7 \hat e^a \otimes \hat e^a \ .
\ee
Our conventions are that $a,b,\ldots$ 
run from 1 to 7 and $\a,\b,\ldots$ from 1 to 6. 
The various manifolds and corresponding vielbeins 
are summarized in the table below.
Since $X$ has $G_2$-holonomy we may assume that the 7-beins
$\hat e^a$ are chosen such that the $\o^{ab}$ are self-dual, 
and hence we know from the above remark that the closed 
and co-closed 3-form $\F$ is simply given by
eq. (\ref{f9a}), i.e.
$\F={1\over 6} \psi_{abc}\, \hat e^a \w \hat e^b \w \hat e^c$.
Although there is such a self-dual choice, in general,
we are not guaranteed
that this choice is compatible with the natural choice of 
7-beins on $X$ consistent with a cohomogeneity-one metric 
as (\ref{f10}). (For any of the three examples cited above, 
the self-dual choice actually
is compatible with a cohomogeneity-one metric.)

\vskip 10.mm
\begin{center}
\begin{tabular}{|c||c|c|c|c|}\hline
{\rm manifolds} & $X$ &  $Y$ &$X_c$ & $X_\l$ \\
\hline
\hline
{\rm vielbeins} & $\hat e^a$ &  $\et^\a$ &$\eb^a$ & $e^a$\\
\hline
{\rm 3-forms} & $\F$ &  &$\phi$ & $\F_\l$ \\
\hline
\end{tabular}
\end{center}
\vskip 5.mm
\centerline{\it Table 1 : The various manifolds,
corresponding vielbeins \ \ \ \  }
\centerline{\it and 3-forms that enter our construction.}
\vskip 5.mm

Now we take the limit $X\to X_c$ in which the 
$G_2$-manifold becomes
exactly a cone on $Y$ so that $\hat e^a\to \eb^a$ with
\be\label{f14}
\eb^\a = r \et^\a \quad , \quad \eb^7 = \d r \ .
\ee
In this limit the cohomogeneity-one metric can be shown to be 
compatible with the self-dual choice of frame 
(see appendix \ref{appcomp}) 
so that we may assume that (\ref{f14}) is such a self-dual 
frame. More precisely, we may assume that the original frame 
$\hat e^a$ on $X$ was chosen in such a way that after taking 
the conical limit the $\eb^a$ are a self-dual frame. 
Then we know that the 3-form $\F$  of $X$ becomes a 
3-form $\f$ of $X_c$ given by the limit of (\ref{f9a}), namely
\be\label{f15}
\f=r^2 \d r \w \xi + r^3 \zeta
\ee
with the 2- and 3-forms on $Y$ defined by
\ba\label{f16}
\xi&=&{1\over 2} \psi_{7\a\b}\, \et^\a \w \et^\b 
\nonumber
\\
\zeta&=&{1\over 6} \psi_{\a\b\g}\,  \et^\a \w \et^\b \w \et^\g \ .
\ea
The dual 4-form is given by
\be\label{f17}
\st\f=r^4 \sty\xi - r^3 \d r \w \sty\zeta
\ee
where $ \sty\xi$ is the dual of $\xi$ in $Y$ (see appendix 
\ref{apphodge} for an explanation of the powers 
of $r$ and the sign). As for the original $\F$, after taking
the conical limit, we still have $\d\f=0$ and $\d\st\f=0$.
This is equivalent to
\ba\label{f18}
\d\xi&=& 3\zeta
\nonumber \\
\d \sty\zeta &=& - 4\sty\xi \ .
\ea
These are properties of appropriate forms on $Y$, and 
they can be checked to be true for any of the three standard $Y$'s.
Actually, these relations show that $Y$ has weak $SU(3)$-holonomy.
Conversely, if $Y$ is a 6-dimensional manifold with weak 
$SU(3)$-holonomy, then we know that these forms exist.
This is analogous to the existence of the 3-form 
$\F_\l$ with $\d\F_\l=4 \l \st\F_\l$ for weak $G_2$-holonomy.
These issues were discussed e.g. in \cite{Hitchin}.
Combining the two relations (\ref{f18}), 
we see that on $Y$ there exists a 2-form $\xi$ obeying
\be\label{f19}
\d\sty\d\xi + 12 \sty\xi=0 \quad , \quad \d\sty\xi=0 \ .
\ee
This equation can equivalently be written as 
$\Delta_Y\xi = 12 \xi$ where 
$\Delta_Y = -\sty\d\sty\d - \d\sty\d\sty$ is the 
Laplace operator on forms on $Y$.
Note that with $\zeta={1\over 3}\d\xi$ we actually have 
$\f=\d\left( {r^3\over 3}\xi\right)$ and $\f$ is 
cohomologically trivial. This was not the case for 
the original $\F$.

We now construct a manifold $X_\l$ with a 3-form $\F_\l$
that is a deformation  of this 3-form $\f$  and
that will satisfy the condition (\ref{f4}) for weak 
$G_2$-holonomy. This general construction is inspired 
by the examples of \cite{Cleyton}. Since weak 
$G_2$-manifolds are Einstein manifolds we need to introduce 
some scale $r_0$ and make the following ansatz for the metric
on $X_\l$
\be\label{f20}
\d s_{X_\l}^2 = \d r^2 +r_0^2 \sin^2 \rh\ \d s_Y^2 \ ,
\ee
with 
\be\label{f21}
\rh = {r\over r_0} \quad , \quad 0\le r \le \pi r_0 \ .
\ee
Clearly, this metric has two conical singularities, one at $r=0$
and the other at $r=\pi r_0$. 
Obviously also, $X_\l$ is a compact manifold since $Y$ is 
compact and $r$ ranges over a closed finite interval with 
the metric remaining finite as $r\to 0$ or $r\to \pi r_0$. 
More specifically, $X_\l$ has finite volume which is 
easily computed to be 
\be\label{f21vol}
{\rm vol}(X_\l) = \int_{X_\l} \sqrt{g} 
= {\rm vol}(Y) \int_0^{\pi r_0} \d r \left( r_0 \sin\rh\right)^6
={5\pi\over 16}\ r_0^7\, {\rm vol}(Y) \ .
\ee
As shown in appendix \ref{app3a}, the metric \ref{f20} 
is ``unique'' in the following sense:
let $Y$ be an Einstein space 
with $\R_{\a\b}=5\delta_{\a\b}$ and let
\be\label{f22}
\d s_{X_\l}^2 = \d r^2 + h^2(r)\ \d s_Y^2 \ .
\ee
Then $X_\l$ is an Einstein space with $\R_{ab}=6\l^2\delta_{ab}$
iff
\be\label{f23}
h^2(r)= r_0^2 \sin^2\left({r\over r_0}\right) 
\quad {\rm with} \quad \l^2=r_0^{-2}
\ee
(up to a possible shift of $r$). This remains true in the limit
$r_0\to \infty$ with $\l=0$ and $h(r)=r$ where one gets 
back the cone.

We see from (\ref{f20}) that we can choose 7-beins $e^a$ on 
$X_\l$ that are 
expressed in terms of the 6-beins $\et^\a$ of $Y$ as
\be\label{f24}
e^\a=r_0 \sin \rh \ \et^\a \quad , \quad e^7=\d r \ .
\ee
Although this is the natural choice, it should be noted
that  it is  {\it not} the one that leads 
to a self-dual spin connection
$\o^{ab}$ that satisfies eq. (\ref{f9}). We know from 
\cite{BDS} that such a self-dual choice of 7-beins must 
exist if the metric
(\ref{f20}) has weak $G_2$-holonomy  
but, as noted earlier, there is no reason why this choice 
should be compatible 
with cohomogeneity-one, i.e choosing $e^7=\d r$. 
Actually, it is easy to see that for weak $G_2$-holonomy, 
$\l\ne 0$, a cohomogeneity-one choice of frame
and self-duality are incompatible:
a cohomogeneity-one choice of frame means $e^7=\d r$ 
and $e^\a=h_{(\a)}(r) \et^\a$ so that 
$\o^{\a\b}={ h_{(\a)}(r) \over h_{(\b)}(r)}
\wt\o^{\a\b}$ and $\o^{\a 7}=h_{(\a)}'(r) \et^\a$. 
But then the self-duality condition for $a=7$ reads 
$\p_{7\a\b}{ h_{(\a)}(r) \over h_{(\b)}(r)} 
\wt\o^{\a\b}=-2\l \d r$. Since $\wt\o^{\a\b}$ 
is the spin connection on $Y$, associated with $\et^\a$, 
it contains no $\d r$-piece, and the self-duality 
condition cannot hold unless $\l=0$. Indeed, 
the examples of self-dual $\o^{ab}$ for weak $G_2$ mentioned 
above were all for cohomogeneity-zero. 

Having defined the 7-beins on $X_\l$ in terms of the 6-beins 
on $Y$, the Hodge duals on $X_\l$ and on $Y$ are 
related accordingly. As shown in appendix \ref{apphodge},
if $\o_p$ is a $p$-form on $Y$, we have 
\ba\label{f27}
\st\left( \d r \w \o_p\right) 
&=& \left( r_0 \sin\rh\right)^{6-2p}
\sty\o_p \nonumber
\\
\st\o_p &=& (-)^p \left( r_0 \sin\rh\right)^{6-2p}
\d r \w \sty\o_p \ ,
\ea
where we denote both the form on $Y$ and its trivial
($r$-independent)
extension onto $X_\l$ by the same symbol $\o_p$.

Finally, we are ready to determine the 3-form $\F_\l$
satisfying $\d\F_\l =\l \st\F_\l$. Inspired by the 
examples considered in \cite{Cleyton}, we make the ansatz
\be\label{f28}
\F_\l= (r_0\sin\rh)^2 \, \d r\w\xi
+ (r_0\sin\rh)^3 \left( \cos\rh\ \zeta + \sin\rh\ \theta\right)
\ .
\ee
Here, the 2-form  $\xi$ and the 3-forms $\zeta$ and $\theta$
are forms on $Y$ which are trivially extended to forms on 
$X_\l$ (no $r$-dependence). 
Note that this $\F_\l$ is not of the form (\ref{f9a}) 
as the last term is not just $\zeta$ but 
$\cos\rh\ \zeta + \sin\rh\ \theta$. 
This was to be expected since the cohomogeneity-one frame 
cannot be self-dual.
The Hodge dual of $\F_\l$ then is  given by
\be\label{f29}
\st\F_\l= (r_0\sin\rh)^4 \sty\xi - (r_0\sin\rh)^3 \, \d r \w
\left( \cos\rh \sty\zeta + \sin\rh \sty\theta\right)
\ee
while
\ba\label{f30}
\d\F_\l&=& (r_0\sin\rh)^2\, \d r\w (-\d\xi+ 3 \zeta)
+ (r_0\sin\rh)^3 \,
\left( \cos\rh\ \d\zeta + \sin\rh\ \d\theta\right)
\nonumber
\\
&+&{4\over r_0} (r_0\sin\rh)^3\, \d r \w  
\left( \cos\rh\ \theta - \sin\rh\ \zeta\right) \ .
\ea
In the last term, the derivative $\partial_r$ has exchanged 
$\cos\rh$ and $\sin\rh$ and this is the reason why both of 
them had to be present in the first place.

Requiring $\d \F_\l = 4\l \st\F_\l$ leads to the following 
conditions
\ba\label{f31}
\d\xi&=& 3\zeta
\\
\label{f32}
\d\theta &=& 4\l r_0 \sty\xi
\\
\label{f33}
\theta &=& -\l r_0 \sty\zeta
\\
\label{f34}
\zeta &=& \l r_0 \sty \theta \ .
\ea
Equations (\ref{f33}) and (\ref{f34}) require
\be\label{f35} 
r_0={1\over \l}
\ee
and $\zeta=\sty\theta\ \Leftrightarrow\ \theta = - \sty\zeta$
(since for a 3-form $\sty(\sty\o_3)=-\o_3$). Then (\ref{f32}) is
$\d\theta=4\sty\xi$, and inserting $\theta=-\sty\zeta$ and 
eq. (\ref{f31}) we get
\be\label{f36}
\d\sty\d\xi + 12 \sty\xi =0 \ .
\ee
But we know from (\ref{f19}) that there is such a two-form 
$\xi$ on $Y$. Then pick such a $\xi$ and let 
$\zeta={1\over 3} \d\xi$ and 
$\theta=-\sty\zeta=-{1\over 3} \sty \d\xi$. We 
conclude that
\be\label{f37}
\F_\l =\left({\sin \l r\over \l}\right)^2 \, \d r\w\xi
+{1\over 3} \left({\sin \l r\over \l}\right)^3
\left(\cos\l r \, \d\xi - \sin\l r \, \sty\d\xi\right)
\ee
satisfies $\d\F_\l = 4\l \st\F_\l$ and that the 
manifold with metric (\ref{f20}) has weak $G_2$-holonomy. 
Thus we have succeeded to construct, for every non-compact 
$G_2$-manifold that is asymptotically (for large $r$) a 
cone on $Y$, a corresponding compact weak $G_2$-manifold $X_\l$
with two conical singularities that look, for small $r$, 
like cones on the same $Y$. Of course, one could start 
directly with any 6-manifold $Y$ of weak $SU(3)$-holonomy.

The quantity $\l$ sets the scale of the weak $G_2$-manifold 
$X_\l$ which has a size of order ${1\over \l}$. As $\l\to 0$, 
$X_\l$ blows up and, within any fixed finite distance from 
$r=0$, it looks like the cone on $Y$ we started with.


\section{Cohomology of the weak $G_2$-manifolds}
\setcounter{equation}{0}

We now want to investigate the cohomology of the weak 
$G_2$-manifolds $X_\l$ we have constructed. We will 
show that it is entirely determined 
by the cohomology of $Y$. 
As before, we always assume that $Y$ is a non-singular 
compact Einstein space with positive curvature.
This implies (see e.g. p. 63 of \cite{Joyce2}) 
that there are no harmonic 
1-forms on $Y$ and $b^1(Y)=0$.

Essentially, we will show that, 
on $X_\l$, harmonic $p$-forms with $p\le 3$ are given by 
those on $Y$, while harmonic $p$-forms with $p\ge 4$ are 
given by their Hodge duals on $X_\l$. In particular, this means that
there are no harmonic 1- or 6-forms on $X_\l$.
 
There are various ways to define harmonic forms which 
are all equivalent on a compact manifold without singularities 
where one can freely integrate by parts. Since $X_\l$ has 
singularities we must be more precise about the definition
we adopt and about the required behaviour of the forms as the 
singularities are approached. 

Physically, when one does a Kaluza-Klein reduction of an 
eleven-dimensional $k$-form $C_k$ one first writes a double
expansion
$C_k=\sum_{p=0}^k \sum_i A^i_{k-p}\w  \f_p^i$
where $A^i_{k-p}$ are $(k-p)$-form fields 
in four dimensions  and the $\f_p^i$ constitute, for each $p$,
a basis of $p$-form fields on $X_\l$. It is convenient to 
expand with respect to a basis of eigenforms of the 
Laplace-operator on $X_\l$. Indeed, the standard kinetic term
for $C_k$ becomes
\ba\label{c1}
\int_{{\cal M}_4\times X_\l} \d C_k \w \st \d C_k
&=&\sum_{p,i}\Bigg(
 \int_{{\cal M}_4} \d A^i_{k-p}\w \st  \d A^i_{k-p} \ \
\int_{X_\l} \f_p^i \w \st\f_p^i 
\nonumber
\\
&+& \int_{{\cal M}_4}  A^i_{k-p}\w \st   A^i_{k-p} \ \
\int_{X_\l} \d\f_p^i \w \st\d\f_p^i \Bigg) \ .
\ea
Then a massless field $A_{k-p}$ in four dimensions arises for
every closed $p$-form $\f_p^i$ on $X_\l$ for which
$\int_{X_\l} \f_p \w \st\f_p$ is finite.
Moreover, the usual gauge condition  $\d\st C_k =0$ leads to
the analogous four-dimensional condition
$\d\st A_{k-p}=0$ provided we also have $\d \st\f_p=0$. 
We are led to the following definition:

\noindent
{\bf Definition :}
An $L^2$-harmonic $p$-form $\f_p$ on $X_\l$ is a $p$-form 
such that
\ba\label{c2}
(i) \quad & &|| \f_p ||^2 \equiv \int_{X_\l}\ \f_p \w \st \f_p < \infty \ , \quad {\rm and}
\\
\label{c3}
(ii) \quad & &\d\f_p =0 \quad {\rm and} \quad \d\st\f_p=0 \ .
\ea
Note that this definition is manifestly invariant under Hodge duality.
We will prove the following proposition.

\noindent
{\bf Proposition :} Let $X_\l$ be a 7-dimensional manifold 
with metric given by (\ref{f20}), (\ref{f21}), i.e.
\be\label{c3a}
\d s_{X_\l}^2 = \d r^2 +r_0^2 \sin^2 \rh\ \d s_Y^2 \ ,\quad
{\rm where} \ \
\rh = {r\over r_0}  , \quad 0\le r \le \pi r_0 \ .
\ee
Then all 
 $L^2$-harmonic $p$-forms $\f_p$ on $X_\l$ for $p\le 3$ 
are given by the trivial ($r$-independent) extensions to $X_\l$ of the 
$L^2$-harmonic $p$-forms $\o_p$ on $Y$. For $p\ge 4$ all
$L^2$-harmonic $p$-forms on $X_\l$ are given by $\st\f_{7-p}$.

\noindent
Since there are no harmonic 1-forms on $Y$ we immediately have the

\noindent
{\bf Corollary :} The Betti numbers on $X_\l$ are given by those 
of $Y$ as
\ba\label{c3b}
b^0(X_\l)=b^7(X_\l)=1\quad &,& \quad
b^1(X_\l)=b^6(X_\l)=0\ , 
\nonumber
\\
b^2(X_\l)=b^5(X_\l)=b^2(Y)\quad &,& \quad
b^3(X_\l)=b^4(X_\l)=b^3(Y) \ .
\ea

Before embarking on the rather lengthy proof, let us make 
some remarks. If we do not require square-integrability 
on $X_\l$, it is obvious that harmonic $p$-forms on 
$Y$ ($p=0, \ldots 6$) carry over to harmonic $p$-forms 
on $X_\l$ as $\f_p=\o_p$. This yields $b^p(Y)$ harmonic
$p$-forms on $X_\l$.
Their Hodge duals on $X_\l$ 
then give another set of harmonic $(7-p)$-forms on $X_\l$ 
as $\f_{7-p}=(-)^p\, h(r)^{6-2p}\, \d r\w \sty\o_p$.
Since $\sty\o_p$ is a harmonic $(6-p)$-form on $Y$, 
there are $b^{6-p}(Y)$ such forms. 
Setting $q=7-p$ ($q=1, \ldots 7$), we see that this 
yields $b^{q-1}(Y)$ harmonic $q$-forms on $X_\l$.
Thus the first set 
gives $b^p(Y)$ and the second set gives $b^{p-1}(Y)$ 
harmonic $p$-forms on $X_\l$.
The Betti numbers then would be 
related as $b^p(X_\l)= b^p(Y)+ b^{p-1}(Y)$. This agrees with
the K\"unneth formula for a space of topology $[0,\pi r_0]\times Y$.
What changes this result is 
the requirement of square-integrability which eliminates 
the harmonic $p$-forms on $X_\l$ for $p=4,5,6$
in the first set, and those with $q=1,2,3$ in the second set.

Another subtle point is the following. 
We also have to show that on $X_\l$ there are no other 
$L^2$-harmonic forms than those that come from the harmonic 
forms on $Y$. This would follow easily from standard arguments
if we could freely integrate by parts with respect to 
$r$. However, due
to the singularities at $r=0$ and $r=\pi r_0$ this is not 
allowed. Instead we have to carefully solve the
``radial'' differential equation and show that there
are no ``square-integrable'' solutions.
Fortunately, the differential equation can be reduced 
to the hypergeometric equation where 
we have explicit formulae for the 
asymptotic behaviour of the solutions at our disposal.

Finally, we remark that it is straightforward to generalise
the proposition to  ($n+1)$- and $n$-dimensional manifolds 
$X$ and $Y$
and/or to more general metrics where $r_0 \sin\rh$ 
is replaced by a
different function $h(r)$ with the same linear asymptotics, so 
that one still has conical singularities. We will give 
the precise statement as a corollary, after having completed 
the proof.

\noindent
{\it Preliminaries :} 
We begin by considering general forms on $Y$, not just harmonic 
ones. For some fixed $p$,
let $\o_p^k$ (resp. $\o^i_{p-1}$) be a basis of $p$-forms 
(resp. $(p-1)$-forms) on $Y$. Since we will only be interested 
in square-integrable forms on $X_\l$, without loss of 
generality,
we will only consider forms on $Y$ that are square-integrable 
on $Y$ with  respect to the standard inner product for $q$-forms
($q=p$ or $p-1$)\footnote{
We are only interested in forms with real coefficients, 
but it is trivial to extend the whole discussion to forms 
with complex coefficients. Then the inner product has to be 
appropriately modified by including complex conjugation 
of the first factor.
}
\be\label{d1}
(\o^k_q \, , \, \o^l_q)=\int_Y \o^k_q \w \sty \o^l_q \ .
\ee
We will assume that the basis is chosen to be
orthonormal with respect to this inner product, which will
simplify the discussion below. Furthermore, we will encounter 
$(\d\o_{p-1}^i\, ,\, \d\o_{p-1}^j)$. Since we can freely 
integrate by parts on the smooth compact manifold $Y$ 
and as $\sty(\sty\o_{p-1})=(-)^{p-1}\o_{p-1}$, we have
\be\label{d2}
(\d\o_{p-1}^i\, ,\, \d\o_{p-1}^j)
=(\o_{p-1}^i\, ,\, -\sty\d\sty\d\o_{p-1}^j) \ .
\ee
Thus $-\sty\d\sty\d$ is a symmetric differential operator on 
forms on $Y$ and we may as well assume that the basis of 
$\o_{p-1}^i$ has been chosen as a basis of orthonormal 
eigenforms of this operator:
\be\label{d3}
-\sty\d\sty\d\ \o_{p-1}^i = \m_i\ \o_{p-1}^i \ ,\quad \m_i\ge 0 \ ,
\ee
so that by (\ref{d2}) we have 
$(\d\o_{p-1}^i\, ,\, \d\o_{p-1}^j) = \m_i\ \delta^{ij}$.

\noindent
{\it Proof :}
To prove the proposition, we consider the most general 
$p$-form $\f_p$ on $X_\l$:
\be\label{d4}
\f_p=\d r\w G_i(r)\, \o^i_{p-1} 
+  g_k(r)\, \o^k_p 
\ee
where sums over repeated indices $i$ or $k$ are understood. 
We want to see which restrictions are imposed on the 
functions $G_i(r)$ and $g_k(r)$ by equations (\ref{c2}) 
and (\ref{c3}) defining an $L^2$-harmonic $p$-form on $X_\l$.

First, $\d\f_p=0$ implies
\be\label{d5}
G_i(r)\, \d\o^i_{p-1} = g'_k(r)\, \o^k_p 
\quad {\rm and}\quad
g_k(r)\, \d\o^k_p  =0 \ .
\ee
We define $\wt F_i(r)=\int_{r^*}^r G_i(\tilde r) \d \tilde r$ 
for some $r^*\in [0,\pi r_0]$. Integrating the first relation
(\ref{d5}) gives
\be\label{d6}
g_k(r) \o_p^k = \wt F_i(r)\,  \d\o^i_{p-1} + \o_p \ ,
\ee
where $\o_p=g_k(r^*) \o_p^k$, and by the second equation (\ref{d5})
we have $\d\o_p=g_k(r^*) \d\o_p^k=0$. 
Hence, $\o_p$ is closed on $Y$ and we can write
$\o_p=\o_p^{(0)} +\d(a_i\o_{p-1}^i)$,
where $\o_p^{(0)}$ is a harmonic $p$-form on $Y$. Finally, if we 
let $F_i(r)=\wt F_i(r)+a_i$, we see from equations
(\ref{d4}) and (\ref{d6}) that
\ba\label{d8}
\f_p&=&\o_p^{(0)} + \d \left( F_i(r) \o^i_{p-1}\right) 
\nonumber
\\
&=&\o_p^{(0)} + F_i(r) \d\o_{p-1}^i 
+ F_i'(r) \d r\w \o_{p-1}^i \ ,
\ea
with $\o_p^{(0)}$ harmonic on $Y$. Note that, in general,
$F_i(r) \o_{p-1}^i$ is not square-integrable so that 
$\f_p-\o_p^{(0)}$ is not $L^2$-exact.

The Hodge dual of such a $\f_p$ on $X_\l$ is 
(recall eq. (\ref{f27}))
\be\label{d9}
\st\f_p=
(-)^p (r_0\sin\rh)^{6-2p} \d r\w
\left( \sty\o_p^{(0)} + F_i(r) \sty\d\o_{p-1}^i \right) 
+ (r_0\sin\rh)^{8-2p} F_i'(r) \sty\o_{p-1}^i \ .
\ee
Then $\d\st\f_p=0$ implies
\ba\label{d10}
\left( (r_0\sin\rh)^{8-2p} F_i'(r)\right)' \sty\o_{p-1}^i 
&=& (-)^p (r_0\sin\rh)^{6-2p}  F_i(r) \d\sty\d\o_{p-1}^i 
\\
\label{d11}
(r_0\sin\rh)^{8-2p} F_i'(r) \d\sty\o_{p-1}^i&=& 0 \ .
\ea
At this point it is useful that we have chosen the $\o_{p-1}^i$
as a basis of eigenforms of the differential operator
$-\sty\d\sty\d$ with eigenvalue $\m_i$ so that eq. (\ref{d10})
becomes
\be\label{d12}
\left( (r_0\sin\rh)^{8-2p} F_i'(r)\right)' =
\m_i (r_0\sin\rh)^{6-2p}  F_i(r) \qquad ({\rm no\ sum\ over\ } i) \ .
\ee

Then every solution to eqs (\ref{d11}) and (\ref{d12}) 
which leads to a square-integrable $\f_p$ yields an 
$L^2$-harmonic $p$-form on $X_\l$. From eqs (\ref{d8})
and (\ref{d9}) we get
\ba\label{d13}
||\f_p||^2&=&
\int_{0}^{\pi r_0}  (r_0\sin\rh)^{6-2p} \d r 
\int_Y \o_p^{(0)} \w\sty  \o_p^{(0)}
\nonumber
\\ 
&+& \sum_i\int_{0}^{\pi r_0} 
\left[  (r_0\sin\rh)^{6-2p} (F_i(r))^2 \m_i
+ (r_0\sin\rh)^{8-2p} (F_i'(r))^2 \right] \d r   \ ,
\ea
where we used eqs (\ref{d2}), (\ref{d3}) and the othonormality 
of the $\o_{p-1}^i$. The cross-terms, linear in $\o_p^{(0)}$ 
dropped out, upon integrating by parts on $Y$.
Since  (\ref{d13}) is a sum of non-negative terms, finiteness of
$||\f_p||^2$ implies finiteness of each of them separately.

In the sum over $i$ we will distinguish the indices 
$i=1, \ldots i_0$ that we will take to correspond to 
$\m_i=0$, from those $i>i_0$ with $\m_i>0$. Since $\m_i=0$ 
means $\d\o_{p-1}^i=0$, eq. (\ref{d8}) can be rewritten as
\ba\label{d14}
\f_p&=&\o_p^{(0)} + \sum_{i=1}^{i_0} F_i'(r)\,  \d r\w \o_{p-1}^i
+ \sum_{i>i_0} \left( F_i'(r)\,  \d r\w \o_{p-1}^i  
+ F_i(r)\,  \d\o_{p-1}^i \right) 
\nonumber
\\ 
&\equiv& \f_p^{(1)} + \f_p^{(2)} + \f_p^{(3)} \ .
\ea

We will show below that if $\f_p^{(3)}\ne 0$ (with $F_i$ 
solutions of (\ref{d12}) with $\m_i>0$) then 
$||\f_p||^2$ diverges. Anticipating this result, we must take 
$\f_p^{(3)}=0$. Then only $F_i(r)$ with $i=1, \ldots i_0$
appear in eq. (\ref{d14}). They are solutions of eq.
(\ref{d12}) with $\m_i=0$ so that
\be\label{d16}
(r_0\sin\rh)^{8-2p} F_i'(r) = c_i \ , \quad i=1, \ldots i_0\ ,
\ee
(with real $c_i$) and from eq. (\ref{d11})
\be\label{d17}
\d\sty\,\wt\o_{p-1}^{(0)} = 0 \quad {\rm with}\quad 
\wt\o_{p-1}^{(0)} = \sum_{i=1}^{i_0} c_i \o_{p-1}^i \ .
\ee
Note that this implies that $\wt\o_{p-1}^{(0)}$ is harmonic on $Y$
and $\st\f_p^{(2)}= \sty\, \wt\o_{p-1}^{(0)}$. Then
\be\label{d18}
||\f_p||^2=
\int_{0}^{\pi r_0} \d r   \left[ (r_0\sin\rh)^{6-2p}
\int_Y \o_p^{(0)} \w\sty  \o_p^{(0)}
+ (r_0\sin\rh)^{2p-8} \sum_{i=1}^{i_0} c_i^2 \right]  \ .
\ee
This converges for $p\le 3$ if and only if $c_i=0$, i.e. 
$\wt\o_{p-1}^{(0)}=0$, and for $p\ge 4$ if and only if 
$\o_p^{(0)}=0$.\footnote{
Note that for $p\ge 4$, 
$\f_p= \sum_{i=1}^{i_0} F_i'(r)\, \d r\w \o_{p-1}^i
= \d \left( \sum_{i=1}^{i_0} F_i(r)\, \w \o_{p-1}^i\right)$
is  square-integrable, but
$\sum_{i=1}^{i_0} F_i(r)\, \w \o_{p-1}^i$ is not, 
confirming the remark below eq. (\ref{d8}).

}
Hence, $\f_p$ is $L^2$-harmonic if and only if
\ba\label{d19}
\f_p&=&\o_p^{(0)} \qquad \qquad {\rm for} \ p\le 3 \ ,
\nonumber
\\
\f_p&=&\st(\sty\,\wt\o_{p-1}^{(0)}) \quad\  \,  {\rm for} \ p\ge 4 \ .
\ea
Since $\sty\,\wt\o_{p-1}^{(0)}$ is a harmonic $(7-p)$-form on 
$Y$, the $L^2$-harmonic $\f_p$ for $p\ge 4$ are exactly the 
Hodge duals of the $L^2$-harmonic $\f_q$ with $q\le 3$.

Finally, we show that in the decomposition (\ref{d14})
of an $L^2$-harmonic $p$-form $\f_p$ the third part 
$\f_p^{(3)}$ must vanish. For this it is enough to show that 
$\f_p^{(3)}$ by itself cannot be $L^2$. In order to do so, 
we assume $F_i(r)\ne 0$ for one or more $i>i_0$ (with $\m_i>0$). 
For these $F_i$, equations (\ref{d11}) and (\ref{d12}) must 
hold. While eq. (\ref{d11}) involves a sum over all $i$, 
eq. (\ref{d12}) holds for every $i$ separately. We want to 
show that there are no (non-vanishing) solutions of 
(\ref{d12}) that lead to a finite norm
\footnote{
The usual argument for this type of problem goes like this:
Assume $F_i$ satisfies eq. (\ref{d12}) with $\m_i>0$.
From the asymptotics of the differential equation one 
immediately deduces that the general solution behaves as
$F_i\sim \a_i\, r^{\n_+}+\b_i\, r^{\n_-}$  as $r\to 0$ with 
$\n_\pm=\tilde p\pm\sqrt{\tilde p^2+\m_i}$ where 
$\tilde p=p-{7\over 2}$, and similarly 
$F_i\sim \g_i\, (\pi r_0-r)^{\n_+}
+\delta_i\, (\pi r_0-r)^{\n_-}$  
as $r\to \pi r_0$. We may choose the coefficients $\a_i$ and 
$\b_i$ but then $\g_i$ and $\delta_i$ are determined through the actual solution of the full differential equation. 
Square-integrability 
(in the sense of eq. (\ref{d13}))
requires $\b_i=0$ and $\delta_i=0$. We may freely choose 
$\b_i=0$, and then determine $\delta_i$. If $\delta_i\ne0$ 
then there is no square-integrable solution. The argument
we present in the text is somewhat different, although 
perfectly equivalent.
}  
of $\f_p^{(3)}$.

To proceed, we now reduce eq. (\ref{d12}) to the 
well-known hypergeometric equation. The required change 
of variables is two-to-one, so first we have to 
reformulate the problem on half the interval, i.e. on 
$[0,{\pi r_0\over 2}]$.
Since eq. (\ref{d12}) is a second-order differential 
equation there 
are two linearly independent solutions. Given the symmetry 
${\cal P}$ that maps $\rh \to \pi - \rh$, one solution can 
be chosen 
${\cal P}$-even and the other ${\cal P}$-odd. The
general solution $F_i$ then is 
$F_i= \a_i F_i^{\rm (even)} + \b_i F_i^{\rm (odd)}$.
When computing $||\f_p^{(3)}||^2$ no cross-terms
$F_i^{\rm (even)} F_i^{\rm (odd)}$ or 
$(F_i^{\rm (even)})' (F_i^{\rm (odd)})'$ survive and
it is enough to show that neither $F_i^{\rm (even)}$
nor $F_i^{\rm (odd)}$  leads to a finite norm. With either choice,
$F_i=F_i^{\rm (even)}$ or $F_i= F_i^{\rm (odd)}$,
the integrand in (\ref{d13}) is ${\cal P}$-even and can
be rewritten as twice the integral over half the interval:
\be\label{d20}
||\f_p||^2\Bigg\vert_{F_i{\rm -\ contribution}} =
2 \int_{0}^{\pi r_0/2} \d r   \left[ (r_0\sin\rh)^{6-2p}
(F_i(r))^2\, \m_i
+ (r_0\sin\rh)^{8-2p}(F_i'(r))^2 \right]  \ ,
\ee
and it is enough to know the solution $F_i$ for 
$0\le r\le {\pi r_0\over 2}$. 
Note that eq. (\ref{d20}) only holds if $F_i$ is the even 
or the odd solution, but not for an arbitrary linear 
combination thereof.
If $F_i$ is ${\cal P}$-even
then $F_i'(\pi r_0/2) =0$ and if $F_i$ is ${\cal P}$-odd
then  $F_i(\pi r_0/2) =0$. Hence we can solve the differential 
equation (\ref{d12}) on the smaller interval $[0,\pi r_0/2]$ 
and select the even or odd $F_i$ through the appropriate 
boundary condition at ${\pi r_0\over 2}\, $.

Now we make the change of variables
\be\label{c27}
x=(\sin\rh)^{-2} \quad , \quad F_i(r)=f(x) \ .
\ee
This is one-to-one on the smaller interval $[0,\pi r_0/2]$:
as $r$ goes from $0$ to $\pi r_0/2$, our new variable $x$ 
goes from $\infty$ to 1.
One has $r_0\partial_r=-2\sqrt{x-1}\, x\, \partial_x$ and
eq.  (\ref{d12}) becomes
\be\label{c28}
4x (x-1) f''(x) + 2\left( (5-2p)(1-x) +1 \right) f'(x) 
- \m_i f(x) = 0 \ ,
\ee
which is the standard hypergeometric equation. 
The vanishing of $F_i(\pi r_0/2)$ translates
into $f(1)=0$, and the vanishing of $F_i'(\pi r_0/2)$ into
the vanishing of $\sqrt{x-1}\, \partial_x f(x)$ at $x=1$. 
Thus we have to find the solutions of the hypergeometric equation
satisfying one or the other of the following boundary conditions: 
\be\label{c29}
{\rm as} \quad x\to 1 \ : \quad 
f(x) \to 0 \quad {\rm or} \quad
\sqrt{x-1}\, \partial_x f(x) \to 0  \ ,
\ee
and show that both solutions lead to an infinite norm of 
$\f_p^{(3)}$.
The contribution of a given solution $f(x)$ to $||\f_p^{(3)}||^2$ is
\be\label{c30}
||\f_p^{(3)}||^2\Bigg\vert_{F_i{\rm -\ contribution}} = r_0^{7-2p}
\int_{1}^{\infty} \left[  f^2(x) \m_i + 4 x (x-1) (f'(x))^2 \right]
{x^{p-4}\over \sqrt{x-1}} \d x\ .
\ee
We can integrate the second term by parts and use the differential
equation. It is precisely such that the  whole
integrand vanishes leaving us only with the boundary terms
\ba\label{c31}
||\f_p^{(3)}||^2\Bigg\vert_{F_i{\rm -\ contribution}} 
&=& 2  r_0^{7-2p} x^{p-3}\ \sqrt{x-1}\, \partial_x f^2(x)
\Big\vert_{x=\infty}
\nonumber
\\
&-&2  r_0^{7-2p} x^{p-3}\ \sqrt{x-1}\, \partial_x f^2(x)
\Big\vert_{x=1} \ .
\ea
But due to the boundary condition (\ref{c29}), only the 
term at $\infty$ remains.
So all we have to do is to determine the two solutions of the 
hypergeometric differential equation (\ref{c28}) that 
satisfy one or the other  boundary condition
 (\ref{c29}) and check 
whether $x^{p-3}\  \sqrt{x-1}\, \partial_x f^2(x)$ has 
a finite limit as $x$ goes to $\infty$. This is done in 
appendix \ref{apphyper} with the result that, for $\m_i>0$,  
$x^{p-3}\  \sqrt{x-1}\, \partial_x f^2(x)
\Big\vert_{x=\infty}$ always diverges.
We conclude that all $F_i$ with $i>i_0$ must vanish, i.e. 
$\f_p^{(3)}=0$. This completes the proof of our proposition.

As already noted, it is straightforward to generalise
the proposition. First, in the metric one can replace 
$r_0 \sin\rh$ by a more general function $h(r)$ provided
it also vanishes linearly in $r$ as the singularities are
approached. Indeed, this is all that was used to decide the
convergence or divergence of the integrals. Furthermore, if
$Y$ is $n$-dimensional, the exponents $6-2p$ and $8-2p$ 
are replaced by $n-2p$ and $n+2-2p$. The only novelty occurs 
for odd $n$. Then, if $p$ is the middle value ${n+1\over 2}$, 
both terms in the equation replacing (\ref{d18}) 
have a factor $(h(r))^{-1}$ and diverge. Also, in this
case, changing variables from $r$ to $\xi$ with 
${\d\xi\over \d r}=(h(r))^{-1}$ and $-\infty<\xi<\infty$,
eq. (\ref{d12}) gets replaced by 
$\partial_\xi^2 F_i(\xi) = \m_i F_i(\xi)$ and the 
r.h.s. of (\ref{d20}) by $\int_{-\infty}^{\infty} \d\xi 
[F_i^2 \m_i + (\partial_\xi F_i)^2]$. Evidently, no 
solution of finite norm exists. We conclude that the 
following generalisation holds:

\noindent
{\bf Corollary :}  Let $Y$ by any smooth compact 
$n$-dimensional manifold and $X$ 
an $(n+1)$-dimensional manifold with metric given by 
\be\label{c32}
\d s^2_X=\d r^2 + h(r)^2\, \d s^2_Y \ ,
\quad r_1\le r\le r_2 \ ,
\ee
with a smooth function $h(r)$ that is non-vanishing
for $r_1< r< r_2$ and behaves as 
$h(r)\sim r-r_1$ as $r\to r_1$ and $h(r)\sim r_2 -r$ 
as $r\to r_2$. We  require that $h(r)$ is 
such that for $\m>0$ the differential equation
\be\label{c33c}
\left( (h(r))^{n+2-2p} f'(r)\right)' =
\m \, (h(r))^{n-2p}  f(r)  
\ee
has no solution $f$ for which
\be\label{c34c}
 \int_{r_1}^{r_2} \d r   \left[ (h(r))^{n-2p}
(f(r))^2\, \m
+ (h(r))^{n+2-2p}(f'(r))^2 \right] 
= (h(r))^{n+2-2p} f(r) f'(r) \Bigg\vert^{r_2}_{r_1}
\ee
is finite.
Then the $L^2$-harmonic $p$-forms 
on $X$ are given, for $p\le {n\over 2}$, by the trivial 
extensions of the harmonic $p$-forms on $Y$, while for 
$p\ge {n+2\over 2}$ they are 
given by their Hodge duals on $X$. If $n$ is odd, 
there are no $L^2$-harmonic 
${n+1\over 2}\,$-forms on $X$.

\section{Conclusions and discussion}

The compactifications of M-theory on manifolds of $G_2$- 
or weak $G_2$-holonomy lead to ${\cal N}=1$ supersymmetry 
in four dimensions. If the compact $G_2$-holonomy manifolds 
have conical singularities,
interesting four-dimensional physics emerges, involving 
charged chiral fermions and anomalies. As a matter of fact, no
explicit metric on such singular spaces exists.
In this paper, we constructed metrics on compact 
seven-manifolds with two conical singularities which carry
weak $G_2$-holonomy. To do so, we started from any
(non-compact) $G_2$-holonomy manifold $X$ which
is asymptotic to a cone on some $Y$ and derived 
various properties of $Y$. The corresponding weak 
$G_2$-holonomy manifolds then can be taken to be 
the direct product of an interval and the 
six-manifold $Y$, where the metric involves a warp 
factor $\sin^2 {r\over r_0}$.
 
Although the compactification on these manifolds
results in a four-dimensional anti-de Sitter space, 
we can read off useful information from models of 
that kind. In particular, we again expect charged chiral 
fermions living at the singularities which will in 
general lead to anomalies. A better understanding 
of the mechanism of anomaly cancellation might be 
obtained by studying these explicitly known weak
$G_2$-manifolds.
 
One question that arose while studying anomalies on 
singular spaces \cite{Witten} was the determination 
of the gauge group $H^2(X;U(1))$ and more generally 
of the cohomology of $X$. For the manifolds $X_\l$
we constructed, these questions were answered 
unambiguously. The reason is that the structure of 
the weak $G_2$-manifold is almost entirely determined 
by the 6-manifold $Y$. In particular, the cohomology of
$X_\l$ can be inferred from the cohomology of $Y$,
with the Betti numbers being related as
$b^p(X_\l)=b^{7-p}(X_\l)=b^p(Y)$ for $p\le 3$. Thus,
Poincar\'e duality and vanishing first Betti number
are maintained although $X_\l$
is singular. An important ingredient was the physically 
motivated restriction to square-integrable forms only.
We also generalised this result to 
arbitrary ${\rm dim}(Y)$ and more general warp 
factors $h^2(r)$. 
As an interesting consequence of these 
facts we find that the gauge group $H^2(X_\l;U(1))$
coincides with $H^2(Y;U(1))$. For the standard examples 
$Y={\bf CP}^3,\ SU(3)/U(1)^2$ and $S^3\times S^3$ we 
thus get gauge groups $U(1),\ U(1)^2$ and no gauge 
group, respectively.

To generate non-abelian gauge groups one could divide
the weak $G_2$-manifolds by $\Gamma_{ADE}$, with
$\Gamma_{ADE}$ acting non-trivially on $Y$. It is rather 
straightforward to discuss the resulting orbifolds 
and what happens at the two conical singularities,
but this is beyond the scope of the present paper.

\vskip 5.mm
\noindent
{\bf\large Acknowledgements}

\noindent
Steffen Metzger gratefully acknowledges support
by the Studienstiftung des deutschen Volkes 
and by the Gottlieb Daimler- und Karl Benz-Stiftung.

\section{Appendix}
\renewcommand{\theequation}{A.\arabic{equation}}
\setcounter{equation}{0}

In this appendix we collect some useful formulas and 
prove some results needed in the main text. 

\subsection{Projectors on the ${\bf 14}$ and ${\bf 7}$ 
of $G_2$\label{appproj}}

Every antisymmetric tensor $A^{ab}$ transforming as 
the {\bf 21} of SO(7) can always be decomposed \cite{BDS} into
a piece $A_+^{ab}$ transforming as the 
{\bf 14} of $G_2$ (called self-dual) and a piece 
$A_-^{ab}$ transforming 
as the {\bf 7} of $G_2$ (called anti-self-dual):
\ba\label{fa0}
A^{ab}&=&A^{ab}_+ + A^{ab}_-
\\
A^{ab}_+&=&{2\over 3} \left( A^{ab}
+{1\over 4} \hat\psi^{abcd} A^{cd}\right) \  ,
\qquad \psi^{abc} A_+^{bc}=0 \ ,
\\
A^{ab}_-&=&{1\over 3} \left( A^{ab}
-{1\over 2} \hat\psi^{abcd} A^{cd}\right) 
={1\over 6} \psi_{abc} \psi_{cde} A^{de} \ ,
\ea
where
\be\label{fa5}
\hat\psi_{abcd}= {1\over 3!}\epsilon^{abcdefg}\psi_{efg}\ .
\ee
In particular, one has
\ba\label{fa0q}
\o^{ab} \g^{ab}= \o_+^{ab} \g_+^{ab} + \o_-^{ab} \g_-^{ab} \ .
\ea

\subsection{Self-duality and the 3-form\label{appsd}}

It will be useful to have an  explicit representation for 
the $\g$-matrices in 7 dimensions. 
A convenient representation is in 
terms of the $\psi_{abc}$ as \cite{BDS}
\be\label{fa3}
(\gamma_a)_{AB} = i(\psi_{aAB} 
+\delta_{aA}\delta_{8B}
-\delta_{aB}\delta_{8A}) \ .
\ee
Here $a=1, \ldots 7$ while $A, B =1, \ldots 8$ and
it is understood that $\psi_{aAB}=0$ if $A$ or $B$ 
equals 8. One then has \cite{BDS}
\ba\label{fa4}
(\gamma_{ab})_{AB} &=& \hat\psi_{abAB}
+\psi_{abA}\delta_{8B}-\psi_{abB}\delta_{8A}
+\delta_{aA}\delta_{bB}-\delta_{aB}\delta_{bA}
\\
(\gamma_{abc})_{AB} &=& 
i\psi_{abc} (\delta_{AB}-2\delta_{8A}\delta_{8B})
-3i\psi_{A[ab}\delta_{c]B}
-3i\psi_{B[ab}\delta_{c]A}
\nonumber
\\
& &-i\hat\psi_{abcA}\delta_{8B}
-i\hat\psi_{abcB}\delta_{8A} \ .
\ea

\noindent
Now we can prove the following:

\noindent
{\bf Lemma :} Let $\F$ be the 3-form that satisfies
$\d\F=0,\ \d \st\F=0$, for $G_2$-holonomy,
or $\d \F=4\l \st \F$, for weak $G_2$-holonomy.\footnote{
Here we do not distinguish between $\F$ and $\F_\l$ but 
use the same symbol $\F$ since both cases are treated 
in parallel.}
Then $\F$ is given by
\be\label{fa1}
\F={1\over 6}\, \psi_{abc}\ e^a\w e^b\w e^c
\ee
if and only if the spin connection is self-dual, 
i.e. satisfies $\psi_{abc} \o^{ab} = -2 \l e^c$.

As explained in \cite{BDS}, from the covariantly
constant spinor (\ref{f1}) one can always construct 
a covariantly constant 3-form
\be\label{fa2}
\F={i\over 6}\,  \eta^T\g_{abc}\eta \ e^a\w e^b\w e^c \ .
\ee
This implies that $\F$ is closed
($\d \F=0$) and co-closed ($\d \st\F=0$).
For weak $G_2$-holonomy, one defines the 3-form
$\F$ in exactly the same way by (\ref{fa2}), but
now $\eta$ obeys the Killing spinor property (\ref{f3}).
It then follows that $\F$ obeys $\d\F=4 \l \st\F$.
Thus (\ref{fa2}) is the correct 3-form $\F$. To see 
under which condition it reduces to (\ref{fa1}) 
we use the explicit representation for 
the $\g$-matrices (\ref{fa3}) given above.  
It is then easy to see that
$\eta^T\g_{abc}\eta \sim \psi_{abc}$ if and only if 
$\eta_A\sim \delta_{8A}$. This means that our 3-form $\F$
is given by (\ref{fa2}) if and only if the covariantly 
constant, resp. Killing spinor $\eta$ only has an eighth 
component, which then must be a constant which we can 
take to be 1. With this normalisation we have
\be\label{fa6}
\eta^T\g_{abc}\eta = -i \psi_{abc} \ ,
\ee 
so that $\F$ is correctly given by (\ref{fa2}).
From the above 
explicit expression for $\g_{ab}$ one then deduces that 
$(\g_{ab})_{AB}\eta_B = \psi_{abA}$ and
$\o^{ab}(\g_{ab})_{AB}\eta_B = \o^{ab} \psi_{abA}$.
Also, $i(\g_c)_{AB}\eta_B = - \delta_{cA}$, so that
eq. (\ref{f1}), resp. (\ref{f3}) reduces to
$\o^{ab}\psi_{abc} =-2\l e^c$.

\subsection{Compatibility of self-duality and 
cohomogeneity-one\label{appcomp}}

Next we want to investigate the compatibility of 
cohomogeneity-one 
and self-dual choices of frame. A cohomogeneity-one choice of 
frame is one where
\be\label{fa7}
e^7=\d r \ , \quad e^\a=h_{(\a)}(r)\  \et^\a
\ee
so that 
\be\label{fa8}
\o^{\a\b}={ h_{(\a)}(r) \over h_{(\b)}(r)}\ \wt\o^{\a\b} \ , 
\quad \o^{\a 7}=h_{(\a)}'(r)\ \et^\a \ .
\ee 
We have already seen in the main text that for weak 
$G_2$-holonomy with $\l\ne 0$ this is incompatible 
with the self-duality condition
$\p_{7\a\b}\o^{\a\b}=-2\l e^7$. So let now $\l=0$.
Then the self-duality conditions become
\ba\label{fa9}
\p_{7\a\b}\ { h_{(\a)}(r) \over h_{(\b)}(r)}\ \wt\o^{\a\b} &=& 0
\nonumber
\\
2 \p_{\a\delta 7}\ h_{(\delta)}'(r)\ \et^\delta 
+ \p_{\a\b\g}\ { h_{(\b)}(r) \over h_{(\g)}(r)}\ \wt\o^{\b\g}  &=& 0
\ .
\ea
In general, this provides constraints on the $h_{(\a)}(r)$ 
since all $r$ dependence is carried by them. One simple 
solution to solve the $r$-dependence in these equations 
is to take all $h_{(\a)}(r)$ equal and $h_{(\a)}'(r)$ to 
be a constant which can be chosen to equal 1. This solution 
of course corresponds to the case where $h_{(\a)}(r)=r$ and
$X$ is a cone on $Y$. Then equations (\ref{fa9}) simply
translate into conditions on the choice of frame on $Y$:
\be\label{fa10}
\p_{7\a\b}  \wt\o^{\a\b} = 0 \ , \quad
\p_{\a\b\g} \wt\o^{\b\g}= - 2  \p_{7\a\b} \et^\b \ .
\ee
These conditions look similar to the self-duality conditions
for weak $G_2$-holonomy, but they are conditions in 
6 dimensions on $Y$. Actually, as pointed out in the main 
text, if $X$ is a cone on $Y$, the $G_2$-holonomy of $X$ implies
that $Y$ has weak $SU(3)$-holonomy \cite{Hitchin}.
Similarly as 
in \cite{BDS} one can show that for a weak $SU(3)$-holonomy 
manifold one can always chose a frame such that (\ref{fa10}) 
holds. This then shows that for a $G_2$-holonomy manifold 
that is a cone on $Y$, the cohomogeneity-one frame can also 
be chosen to be self-dual.

\subsection{Relating Hodge duals\label{apphodge}}

We need to relate Hodge duals on the 7-manifolds $X$, 
$X_c$ or $X_\l$ to the Hodge duals on the 6-manifold $Y$. 
For the present purpose, we do not need to specify the 
7-manifold and just call it $X_7$. We assume that the 
7-beins of $X_7$, called $e^a$, and the 6-beins of $Y$, 
called $\et^\a$ can be related by
\be\label{fa20}
e^7=\d r \ , \quad e^\a= h(r) \et^\a \ .
\ee
We denote the Hodge dual of a form $\pi$ on $X_7$ simply
by $\st\pi$ while the 6-dimensional Hodge dual of
a form $\sigma$ on $Y$ is denoted $\sty\sigma$.

The duals of $p$-forms on $X_7$ and on $Y$ are defined in 
terms of their respective vielbein basis, namely
\be\label{fa21}
\st\left( e^{a_1} \w \ldots \w e^{a_p}\right) ={1\over (7-p)!}
\e^{a_1\ldots a_p}_{\phantom{a_1\ldots a_p} b_1\ldots b_{7-p}}
e^{b_1} \w \ldots \w e^{b_{7-p}} 
\ee
and
\be\label{fa22}
\sty\left( \et^{\a_1} \w \ldots \w \et^{\a_p}\right) ={1\over (6-p)!}
\e^{\a_1\ldots \a_p}_{\phantom{\a_1\ldots \a_p} \b_1\ldots \b_{6-p}}
\et^{\b_1} \w \ldots \w \et^{\b_{6-p}} \ .
\ee
Here the $\e$-tensors are the ``flat'' ones that equal $\pm 1$.
Expressing the $e^a$ in terms of the $\et^\a$ provides the 
desired relation. In particular,
for a $p$-form $\o_p$ on $Y$ we have 
\ba\label{fa23}
\st\left( \d r \w \o_p\right) &=& h(r)^{6-2p}
\sty\o_p \nonumber
\\
\st\o_p &=& (-)^p h(r)^{6-2p}
\d r \w \sty\o_p \ ,
\ea
where we denote both the form on $Y$ and its trivial
extension onto $X_\l$ by the same symbol $\o_p$.

\subsection{Curvature in $(n+1)$ dimensions from 
curvature in $n$ dimensions \label{app3a}}

Suppose that the metric $\d s_X^2$ on an $(n+1)$-dimensional
manifold $X$ is given in terms of the metric $\d s_Y^2$ on an 
$n$-dimensional manifold $Y$ by
\be\label{r1}
\d s_X^2 = \d r^2 + h(r)^2\,  \d s_Y^2 \ .
\ee
Similarly as before, denote vielbeins on $Y$ by $\et^\a$, 
$\a=1, \ldots n$ and those on $X$ by $e^a$, $a=0,1,\ldots n$
so that $e^0=\d r$ and $e^\a=h(r)\, \et^\a$. Denote the 
spin-connections by $\wt\o^{\a\b}$ and $\o^{ab}$ respectively. 
Then
\be\label{r2}
\o^{\a\b}=\wt\o^{\a\b} \qquad , \qquad 
\o^{\a 0} = {h'\over h} e^\a \ .
\ee
It follows that the curvature 2-forms $R^{ab}$ of $X$ 
and $\wt R^{\a\b}$ of $Y$ are
related as
\be\label{r3}
R^{\a\b} = \wt R^{\a\b} -\left({h'\over h}\right)^2 e^\a\w e^\b
\qquad , \qquad
R^{\a 0} = -{h''\over h}\, e^\a\w e^0 \ .
\ee
Note that this implies that the components of the curvature tensors
are related as
\be\label{r4}
R^{\a\b}_{\ \ \ \g\dd} = {1\over h^2} \wt R^{\a\b}_{\ \ \ \g\dd}
 -\left({h'\over h}\right)^2 
\left( \dd^\a_\g \dd^\b_\dd - \dd^\a_\dd \dd^\b_\g \right)
\qquad , \qquad
R^{\a 0}_{\ \ \ \g 0} = -{h''\over h}\, \dd^\a_\g \ .
\ee
This yields for the Ricci tensor
\be\label{r5}
{\cal R}^\a_{\ \g}
= {1\over h^2} \left( \wt{\cal R}^\a_{\ \g}
-(n-1)\, h'^2\, \dd^\a_\g - h h''\, \dd^\a_\g \right)
\qquad , \qquad
{\cal R}^0_{\ 0} 
= {1\over h^2} \left( -nh h''\right)
\ee
and ${\cal R}^0_{\ \a} = {\cal R}^\a_{\ 0}=0$.
Now suppose  $Y$ is an Einstein space with 
$\wt{\cal R}^\a_{\ \g}=\m^2\, \dd^\a_\g$ ($\m$ is real or imaginary,
and necessarily vanishes if $n=1$) so that
\be\label{r6}
{\cal R}^\a_{\ \g} = {1\over h^2} 
\left( \m^2 -(n-1)\, h'^2  - h h'' \right) \dd^\a_\g \ .
\ee
Then $X$ is an Einstein space with ${\cal R}^a_{\ c}=n\l^2\, \dd^a_c$
($\l$ real or imaginary) if and only if
\be\label{r7}
-{h''\over h}=\l^2 \quad {\rm and} \quad 
(n-1)\, h'^2 = \m^2 - (n-1)\, \l^2\, h^2 \ .
\ee
The first equation is solved by
\be\label{r8}
h(r) = r_0 \sin \l(r-r_1)
\ee
and then, for $n\ge 2$, the second equation implies
\be\label{r9}
r_0^2\, \l^2= {\m^2\over n-1} \ .
\ee
For $n=6$ and $\m^2=5$ we get eq. (\ref{f23}).

\subsection{Properties of the hypergeometric 
equation\label{apphyper}}

Here, we want to show that all 
solutions $f(x)$ of the 
hypergeometric equation (\ref{c28}) with $\m_i>0$ 
that satisfy either of the two boundary conditions 
(\ref{c29}) are such that 
$x^{p-3}\ \sqrt{x-1}\, \partial_x f_i^2(x)$ diverges 
as $x\to\infty$.

The standard form of the hypergeometric equation is 
\cite{Bateman}
\be\label{hyper}
x(1-x) f''(x) + [c-(a+b+1)x] f'(x) - ab f(x) = 0 \ .
\ee
The coefficients of our hypergeometric equation
(\ref{c28}) correspond to 
\be\label{c31a}
a={1\over 2} \left( \tilde p + \sqrt{\tilde p^2+\m_i}\right)\ ,
\quad
b={1\over 2} \left( \tilde p - \sqrt{\tilde p^2+\m_i}\right)\ ,
\quad c=p-3 \ ,
\ee 
with $\tilde p\equiv p-{7\over 2}$. 
As is well-known \cite{Bateman}, there is a 
systematic way to obtain pairs of linearly independent 
solutions of the hypergeometric differential equation with  
cuts on  various combinations of the intervals 
$(-\infty,0]$, $[0,1]$ and $[1,\infty)$. They are given 
by the Kummer solutions $u_1, \ldots u_6$ each of which 
can be expressed in four different ways. Since we 
implicitly assumed that the $F_i(r)$ are real, we have to be 
careful to choose  solutions $f(x)$ that are real on the 
positive real axis for $x\ge 1$.

For our present purpose, it is convenient to consider 
the following pair of linearly independent solutions
\ba\label{c32f}
u_2(x)&=&F(a,b;a+b+1-c;1-x) \ ,
\\
u_6(x)&=&(1-x)^{c-a-b} F(c-a,c-b;c+1-a-b;1-x) \ ,
\ea
where $a,b,c$ are given above and $F$ is the usual 
hypergeometric function. The function $F$ has a cut when its 
argument $1-x$ is in the interval $[1,\infty)$. But this 
corresponds to $x<0$ and is irrelevant to us. However, 
for $u_6$,
the power of $1-x$ in front of $F$ has a simple square-root 
branch cut from 1 to $+\infty$ since $c-a-b={1\over 2}$.
Hence we define
\ba\label{c33}
f_1(x)&=&u_2(x)
\nonumber
\\
f_2(x)&=& e^{i\pi(c-a-b)}  u_6(x+i\e)=
 (x-1)^{c-a-b} F(c-a,c-b;c+1-a-b;1-x)
\nonumber
\\
\ea
which are real and defined without any ambiguity for 
real $x\ge 1$ and
satisfy the boundary condition (\ref{c29})
since 
\ba\label{c33a}
f_1(1)&=&1 \quad , \quad 
2\sqrt{x-1} \partial_x f_1(x) \Big\vert_{x=1} = 0 \ ,
\nonumber
\\ 
f_2(1)&=&0 \quad , \quad 
2\sqrt{x-1} \partial_x f_2(x) \Big\vert_{x=1} = 1 \ .
\ea
To investigate the behaviour of these
solutions as $x\to\infty$, one uses the formula 
\cite{Bateman} for the analytic continuation expressing 
$u_2$ or $u_6$ as  linear combinations of $u_3$ and $u_4$, 
both of 
which behave as powers of $x$ as $x\to\infty$:
\ba\label{c34}
f_1(x) &\sim& 
{\Gamma(a+b+1-c)\Gamma(b-a)\over \Gamma(b+1-c)\Gamma(b)} 
\left(x^{-a} + \ldots\right)
+
{\Gamma(a+b+1-c)\Gamma(a-b)\over \Gamma(a+1-c)\Gamma(a)} 
\left(x^{-b} + \ldots\right) \ ,
\nonumber
\\
f_2(x) &\sim& 
{\Gamma(c+1-a-b)\Gamma(b-a)\over \Gamma(1-a)\Gamma(c-a)} 
\left(x^{-a} + \ldots\right)
+
{\Gamma(c+1-a-b)\Gamma(a-b)\over \Gamma(1-b)\Gamma(c-b)} 
\left(x^{-b} + \ldots\right) \ ,
\nonumber
\\
\ea
where the dots indicate terms that are subleading by 
integer powers of $x$. In our case, $-b>-a$ and the 
leading terms in $f_1$ and $f_2$ as $x\to\infty$ 
are the $x^{-b}$-terms, unless their
coefficients, which we will call $d_1$ and $d_2$, happen to 
vanish. This can only happen if $a+1-c$, $\ a$,
$\ 1-b$ or $c-b$ is a non-positive integer, which is 
impossible since $\sqrt{\tilde p^2+\m_i} > \tilde p$. 
We conclude that, as $x\to\infty$, for both solutions we have 
\be\label{c35}
x^{p-3}\ \sqrt{x-1}\, \partial_x f_i^2(x)
\sim d_i^2\ x^{\tilde p -2b} 
= d_i^2\ x^{\sqrt{\tilde p^2+\m_i} }  \quad
\quad {\rm as}\ x\to\infty \quad  , \quad i=1,2\ .
\ee
This diverges for all $p$ and $\m_i>0$ and, 
by eq. (\ref{c31}), 
so does the corresponding contribution to
$||\f_p^{(3)}||^2$.



\end{document}